\documentclass{INTERSPEECH2023}

\interspeechcameraready

\title{The Power of Prosody and Prosody of Power: An Acoustic Analysis of Finnish Parliamentary Speech}
\name{Martti Vainio, Antti Suni, Juraj Šimko, Sofoklis Kakouros}
\address{
  University of Helsinki, Finland}
\email{\{martti.vainio,antti.suni,juraj.simko,sofoklis.kakouros\}@helsinki.fi}

\begin{document}

\maketitle
 
\begin{abstract}

\noindent 

\noindent Parliamentary recordings provide a rich source of data for studying how politicians use speech to convey their messages and influence their audience. This provides a unique context for studying how politicians use speech, especially prosody, to achieve their goals. Here we analyzed a corpus of parliamentary speeches in the Finnish parliament between the years 2008-2020 and highlight methodological considerations related to the robustness of signal based features with respect to varying recording conditions and corpus design. We also present results of long term changes pertaining to speakers' status with respect to their party being in government or in opposition. Looking at large scale averages of fundamental frequency - a robust prosodic feature - we found systematic changes in speech prosody with respect opposition status and the election term. Reflecting a different level of urgency, members of the parliament have higher $f_0$ at the beginning of the term or when they are in opposition.


\end{abstract}

\section{Introduction}

Parliamentary speech is a popular topic of inquiry, but the vast majority of research has relied on methods applied to speech transcriptions.  The opposition--government dynamics has been widely studied using sentiment analysis identifying a gap between the sentiment expressed by opposition and government representatives, with the opposition being often more negative than the government  \cite{abercrombie2020sentiment,proksch2019multilingual}.
Despite the importance of prosody in parliamentary speech, there has been limited research into this area (see, e.g., \cite{karpinski2022high}). 
 This is surprising given that prosody plays a crucial role in conveying meaning, emotions, and attitudes in speech.

Although parliamentary speech is formal in style, it is not free of emotional coloration. The speakers are likely to manifest their attitudes and express different levels of arousal (negative or positive) during their turns. In addition to their political views with respect to the discussed topic, these attitudes might be modulated by their current political status in terms of representing the ruling government or the opposition. And these attitudes can be expected to be communicated through acoustic-prosodic features of speech. Prosodic analysis can thus serve as a basis of studying social phenomena like affective polarization \cite{iyengar2012affect}, i.e., emotional distance between individuals or groups with opposing political views.





The performance of data-driven ML systems depends on the level of consistency of the data used for training. These can affect the acoustic properties of the speech signals and therefore the extracted signal-based features in systematic ways. For example, different microphones, recording environments, and recording devices can introduce confounds which can lead to superficially overblown classification results, and therefore misleading conclusions.

 This paper addresses these potential signal-based signatures of parliamentary speaking turns based on the government-opposition dichotomy in the Finnish parliamentary speech using a Finnish parliament corpus \cite{virkkunen2022finnish}. We report the performance of classifiers operating on representations obtained by self-supervised learning (SSL). 
We supplement this ML based approach by an analysis of traditional signal-based prosodic features in terms of their robustness to the time-dependent variability, and select $f_0$ as a relatively reliable proxy of features related to emotional arousal. We provide a statistical analysis of this easily interpretable characteristic of speech prosody with respect to political status of the members of parliament (MPs).

 



The Finnish Parliament is composed of a single chamber with 200 members, who are elected every four years through a proportional representation system. The government is formed after negotiations lead by a prime minister candidate from a party who won the elections. The process normally results in a formation of a coalition government, where government parties hold the majority of seats in the parliament. As a norm, two of the three traditional major parties 
 are in a government and one in opposition. That means that there is a relatively large number of longer serving MPs that, over the years, belonged to both investigated status categories, allowing for a longitudinal study on a controlled, stable subset of the parliamentary data.



\section{The corpus and data processing}
In the next we describe the data used in our experiments as well as the feature extraction and post-processing of the features.
\subsection{Finnish parliament corpus}
While all Finnish parliamentary sessions are available on-line\footnote{verkkolahetys.eduskunta.fi,  speaker diarizations and transcriptions available from year 2019 onwards}, for convenience, for the current study we utilized the previously collected Finnish parliament ASR corpus \cite{virkkunen2022finnish}. The full corpus consists of over 3000 hours of speech from 449 speakers, recorded from 743 plenary meetings between the years 2008-2020. The speeches have been automatically split to short utterances, mostly under 15 seconds in length. In addition to transcriptions for ASR training, both speaker Id and recording year are included in the corpus. Using public sources, we enriched the annotations with speakers' party affiliation, sex, age (at the time of recording), and opposition/government status (i.e., whether the given utterance was given by the MP when representing opposition or the ruling coalition). In order to eliminate the variability of MPs between different election terms, we selected a subset of the corpus consisting of speech from only those MPs who had served for at least three election terms, who had served in both opposition and government role, and who had speech material recorded from all of their years of service.  The resulting subset consisted of approximately 900 hours or 402,617 utterances of speech from 57 MPs (25 female, 32 male) with 6 different party affiliations, covering the political spectrum. 

\subsection{Feature extraction}

We extract a standard set of acoustic features using eGeMAPS \cite{eyben2015geneva}. The feature set includes 88 different features and feature functionals including means and standard deviations of $f_0$, loudness, spectral tilt, and MFCCs. In our subsequent analysis we only use a small subset (see Tab.\ref{tab:r-squares} including the following functionals over each utterance: mean and standard deviation of $f_0$, mean loudness, mean MFCC1, and the mean number of voiced frames per second. 

To further evaluate potential separability of the acoustic data in terms of MPs' government--opposition status, we extracted state-of-the-art (SOTA) self-supervised representations from speech. Self-supervised learning (SSL) has been driving SOTA results in many areas of speech technology, including areas where prosody is in question. Representations from SSL models are many orders of magnitude higher in dimensionality than standard acoustic features such as MFCCs, thus, their representation space, and their potential to capture relevant features in speech is much higher. We are extracting SSL representations using XLS-R, a large-scale model for cross-lingual speech representation learning based on wav2vec 2.0 \cite{babu2021xls}. We selected XLS-R as it includes approximately 14,000 hours of training data from Finnish speech, including recordings from VoxPopuli, a corpus of parliamentary speech from the European parliament \cite{wang2021voxpopuli}.

The features were extracted from the pre-trained XLS-R by taking the representation from the last transformer layer of the model. To obtain an utterance-level description, the feature-level representations are pooled by taking the mean of the features over each utterance. This results into a $1024$ dimensional vector for each utterance in our data.

\section{ML classification approach}

The XLS-R features were used to train a feed-forward DNN classifier with government and opposition speech as the target classes. In order to address the possible speaker dependence of the classification performance, the classifiers were trained on two types of data splits, speaker-dependent (SD) and speaker-independent (SI) ones. 
For the SD splits, the data was sampled uniformly, with 85\%-15\% train-test split. For SI splits (SI), the data of four MPs (2 male and 2 females, with most balanced  opposition / coalition utterance counts) were left out from training set and used as a test set. 
For each type of data split we generated three different sets (i.e., a total of 6 experiments) and report the average classification results in Tab.~\ref{tab:classification-results}. The table also contains two types of baseline, a simple majority vote classification (MAJ), with each utterance classified in the majority class (government in this case), and speaker dependent majority vote (MAJ-S), where the utterance is classified in a majority class for the given speaker (as belonging to the class for which the database contains more samples for the given MP).

A feed-forward network with four hidden layers was used for training. The overall network layout is the following $L = [256, 128, 64, 32]$, where $L$ is the network layer. We trained the network for 100 epochs with a batch size 200 and learning rate of $lr$=1$e$-4. During training, we applied dropout regularization to the network to prevent overfitting. Specifically, we applied a dropout rate of $p$=0.1 to all layers except the last hidden layer. To optimize the network we used a binary cross-entropy loss function and adam optimizer.

\begingroup
\setlength{\tabcolsep}{4pt}
\begin{table}[th]
  \caption{Classification results (\% mean and std) with XLS-R and DNN for speaker-dependent (SD) and speaker-independent (SI) splits. Results are averaged over runs.}
  \label{tab:classification-results}
  \centering
  \begin{tabular}{l c c c c}
    \toprule
    & \textbf{$ACC$} & \textbf{$PRC$} & \textbf{$RCL$} & \textbf{$F1$}
 \\
    \midrule
SD   & 76.4 (4.1) & 73.9 (8.0) & 74.6 (5.3) & 73.9 (2.3)\\
SI   & 70.1 (6.5) & 65.0 (24.9) & 58.9 (2.0) & 60.2 (12.3)\\
    \midrule
MAJ-S   & 74.4 & 80.3 & 75.1 & 77.6\\
MAJ   & 55.4 &  &  & \\
    \bottomrule
  \end{tabular}
\end{table}
\endgroup

The classifiers' performance on the speaker dependent (SD) splits shows that, given speaker information, the approach using XLS-R features can achieve promising results, outperforming the relevant majority baseline (MAJ-S). For SI data split, the performance drops below the MAJ-S baseline but remains above speaker-agnostic majority one. These results are compatible with possible a presence of prosodic differences between MP speaker styles in their different political positions.

The difference in terms of performance of the two splits indicates a relatively strong dependence of represented speaker-based prosodic characteristics. The performance measures for the data split with unseen speakers (SI), particularly the high standard deviation of precision measure, also point at a high dependence of classifying performance on speaker idiosyncrasies, namely on speakers selected in the test set (this could of be mitigated by an increase in the number of cross-validation iterations; currently only three). 

\section{Prosodic features}




In a longitudinal corpus as such analysed here one can expect changes in recording conditions systematically varying with replacements of the old recording technologies, changes in seating arrangements of the parliament, microphone distance, room acoustics, or signal processing effects (e.g., compression) that has possibly been applied to the signals during or after the recording. This type of temporal variation can have detrimental effects on the interpretability of classification results such as reported above; the classifiers can potentially learn systematic regularities of interaction between the changes in recording conditions potentially coinciding with changing political status of the individual MPs.

\begin{figure}[t]
   \centering
   \includegraphics[width=1.0\linewidth]{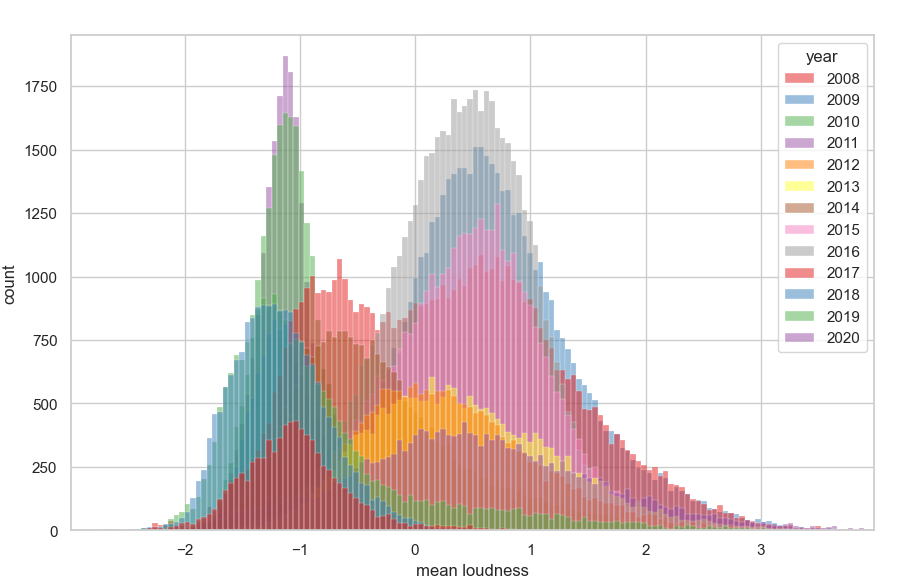}
   \caption{Histogram of speaker-normalized loudness colored by year}
   \label{fig:loudness-hist}
 \end{figure}

Fig.~\ref{fig:loudness-hist} with the yearly distributions of speaker-normalized loudness measure shows presents an example of a strong dependence of the values on the time of recording, presumably reflecting the recording conditions rather than a genuine changes in this prosodic characteristics.

In order to evaluate the degree of this dependence statistically 
we fitted linear models with the year of recording (treated as a factor) as the predictors and the selected eGeMAPS features as dependent variables.
Tab.~\ref{tab:r-squares} lists adjusted $r$-square values for these fits. The higher the adjusted $r$-square (the variance explained the model), the greater the presumed dependence of the signal-based feature on the yearly variations.

\begin{table}[h]
  \caption{Adjusted r-square values for linear models predicting the given feature from \textit{year} information alone.}
  \label{tab:r-squares}
  \centering
  \begin{tabular}{l  r }
    \toprule
    \textbf{feature} & \textbf{adj.~r-square}
 \\
    \midrule
    mean $f_0$ & 0.006     \\
    voiced per sec  & 0.015   \\
    std $f_0$  & 0.064            \\
    mean MFCC1  & 0.120              \\
    mean loudness  & 0.404       \\
    \bottomrule
  \end{tabular}
  
\end{table}

As expected, the features with less dependency on detailed spectral characteristics of the signal (mean and std $f_0$, proportion of voicing) show less systematic dependence compared to spectral features (MFCC) and the loudness ones.

In what follows we will concentrate on the feature least depended on the year of recording, namely the mean $f_0$.

\subsection{Statistical analysis of mean $f_0$ feature}





\begin{figure}[h]
  \centering
  \includegraphics[width=\linewidth]{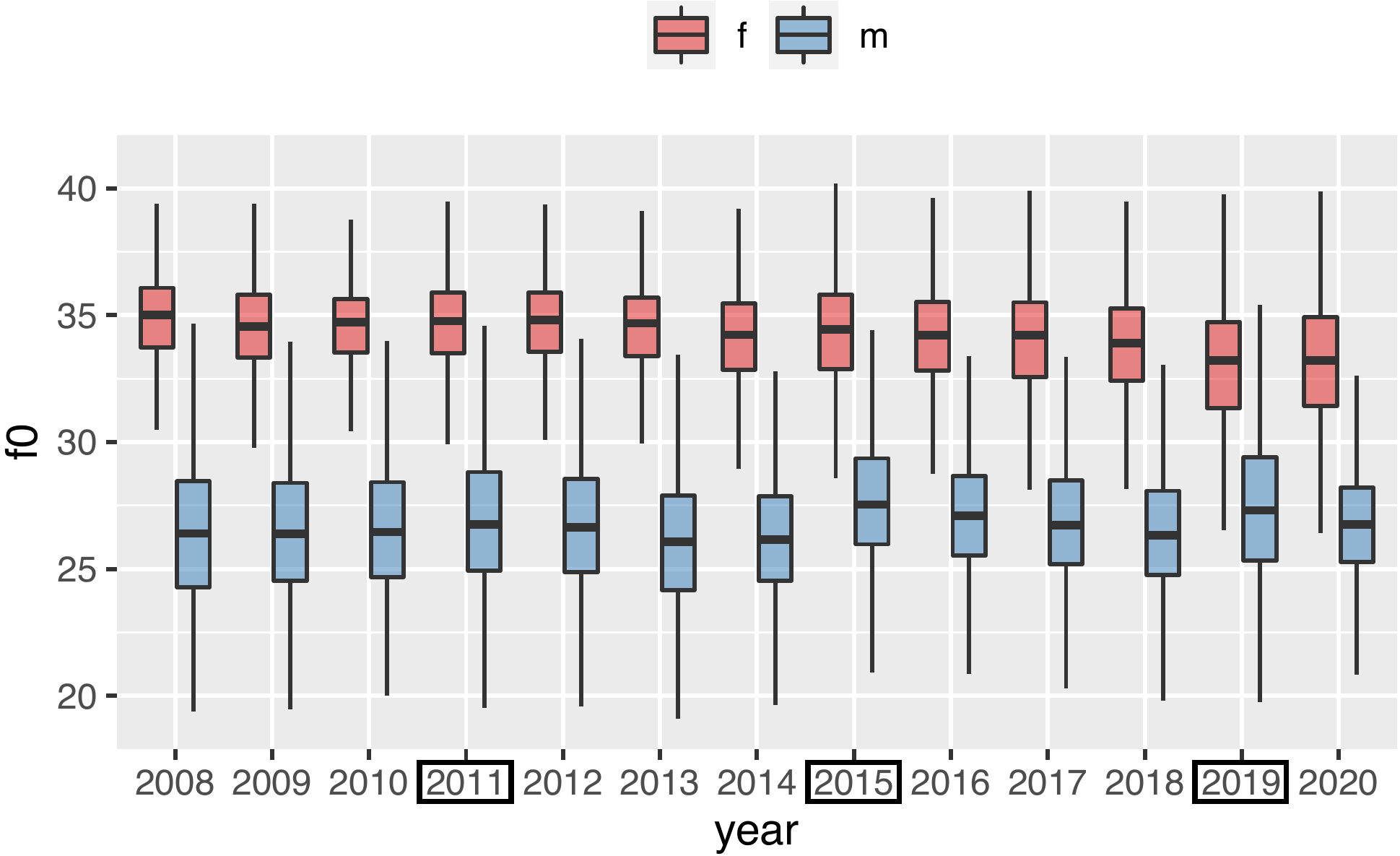}
  \caption{Mean $f_0$ in semitones relative to 27.5 Hz for female and male speakers throughout the years. Election years are marked on the year axis.}
  \label{fig:year_sex}
\end{figure}

While relatively robust with respect to recording conditions, utterance-level mean pitch is obviously highly speaker dependent. The main source of variation is related to the size of the speaker's vocal tract, manifested, for example, in the difference between male and female speakers. There are also other factors influencing $f_0$, especially if any longitudinal phenomena are studied. For instance, with respect to fundamental frequency, it is known that the average level changes during a speakers lifetime.  Moreover, the changes are different depending on the speakers' biological sex.  Typically for females the average $f_0$ falls linearly with age -- for men the changes follow a falling-rising pattern until fairly late in life \cite{hollien1972speaking, harnsberger2008speaking}. 

Fig.~\ref{fig:year_sex} shows the average fundamental frequency values for all speakers by year and sex.  The visible convergence of male and female values broadly follows the above mentioned age related changes, where female speech becomes lower in pitch and male speech higher during adulthood.
In addition, concentrating on election terms, the figure shows a somewhat higher average pitch values for the election years (when a new parliament gets inaugurated), followed by a downtrend, in particular for the male MPs. We will return to these observation later in this section.

In order to mitigate the speaker dependencies we normalized the $f_0$ mean values by subtracting median value for each speaker from the speaker's mean $f_0$ values. We further balanced the data by keeping an equal number of opposition and government utterances for each individual speaker (1150 for each political position; this number maximizes the size of the final subset). This subsetting was possible for 32 speakers (15 female)--for the others the corpus did not contain enough utterances in at least one of the positions--yielding altogether 73,600 utterances used for the subsequent statistical analyses.

\begin{figure*}[t]
  \centering
  \includegraphics[width=\textwidth]{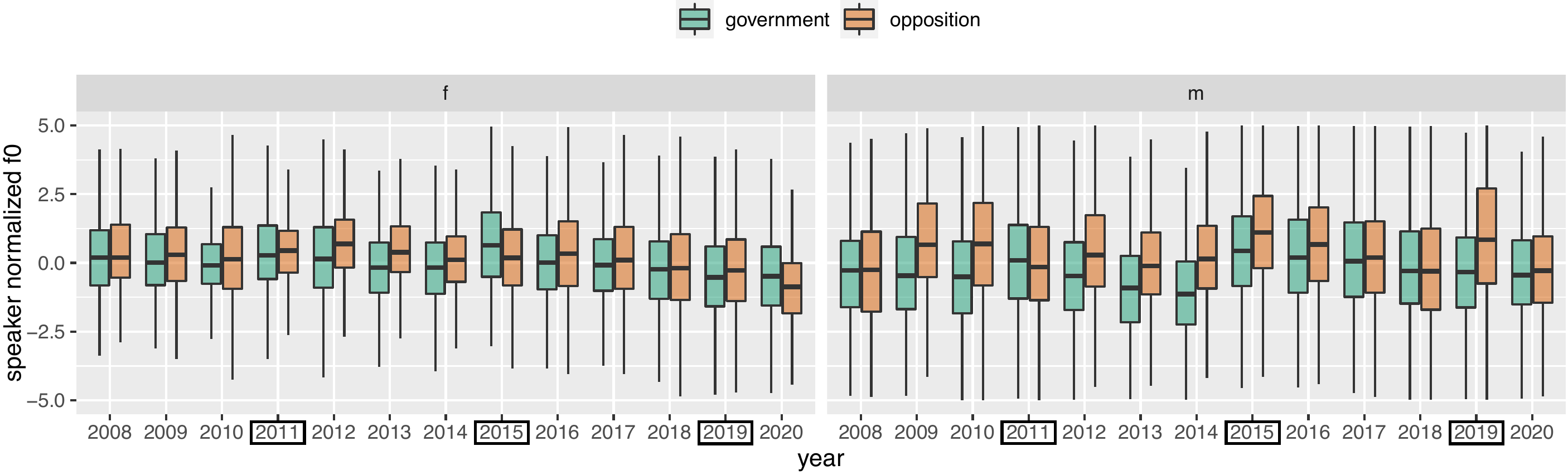}
  \caption{Normalized mean $f_0$ representing government and opposition throughout the years, separately for female and male speakers.}
  \label{fig:year_sex_status}
\end{figure*}

Fig.~\ref{fig:year_sex_status} shows normalized mean $f_0$ averages per year divided by the current (for the given year) political affiliation (government or opposition) of the MPs, separately for females and males.

As the figure indicates, the average normalized $f_0$ was in most cases higher for opposition speakers compared to the ones representing the government.  To test this observation statistically, we fitted a mixed effect model with the normalized $f_0$ as the dependent variable, political affiliation and speaker's sex (and their interaction) as the fixed dependent variables and the year of recording (treated as a factor) as a random effect. The choice of the random effect is to account for possible left-over yearly variation associated with, for example, the noted long-term declination of average pitch values for the female speakers.


\begin{table}[th]
  \caption{Fixed effect coefficients of the regression of status and sex against mean $f_0$.}
  \label{tab:f0_sex_status}
  \centering
  \begin{tabular}{l c c c}
    \toprule
    & \textbf{Estimate} & \textbf{Std. Error}  & \textbf{$t$-value}
 \\
    \midrule
(Intercept)       &     -0.07  &  0.08 & -0.91 \\
sex\_male              &     -0.07  &  0.02 & 3.43 \\
status\_opp   &    0.21  &  0.02 & 9.11 \\
sex\_male:status\_opp & 0.40  &  0.03 & 12.21 \\
    \bottomrule
  \end{tabular}
\end{table}

The effect of the intercept (corresponding to the $f_0$ mean estimate for females speaking as government MPs) is not significant\footnote{For the random effects models with a large number of observations, significance level expressed with $p$-value $=0.05$ roughly corresponds to the absolute value of $t$-value equal to 2; all effects with higher absolute value can be considered significant.}. The normalized mean $f_0$ for male speakers speaking on behalf of the government is significantly lower than for females with the same political status ($t$-value = 3.43). More interestingly, the females use significantly higher $f_0$ when representing opposition than when speaking on behalf of the government (status\_opp(osition) effect; $t$-value = 9.11). As indicated by the interaction estimate, the difference between opposition and coalition is significantly even more pronounced for the male speakers compared to the females ($t$-value = 12.21). In short, MPs of both sexes use significantly higher pitch when in opposition, males significantly more so than females.

Returning to the observation of the long-term sex based trends in mean $f_0$, we have fitted a mixed effect linear regression models with the normalized mean $f_0$ as a dependent variable and the year of recording as a predictor (treated as a continuous variable) and speaker as a random factor, separately for male and female politicians. As observed, the fixed effect of recording year is significantly negative for females ($t$-value = -20.58) and significantly positive for males ($t$-value = 5.78). The effect size is about three-times greater for females than for males (-0.064 \textit{vs.} 0.022).













\begin{figure}[t]
  \centering
  \includegraphics[width=\linewidth]{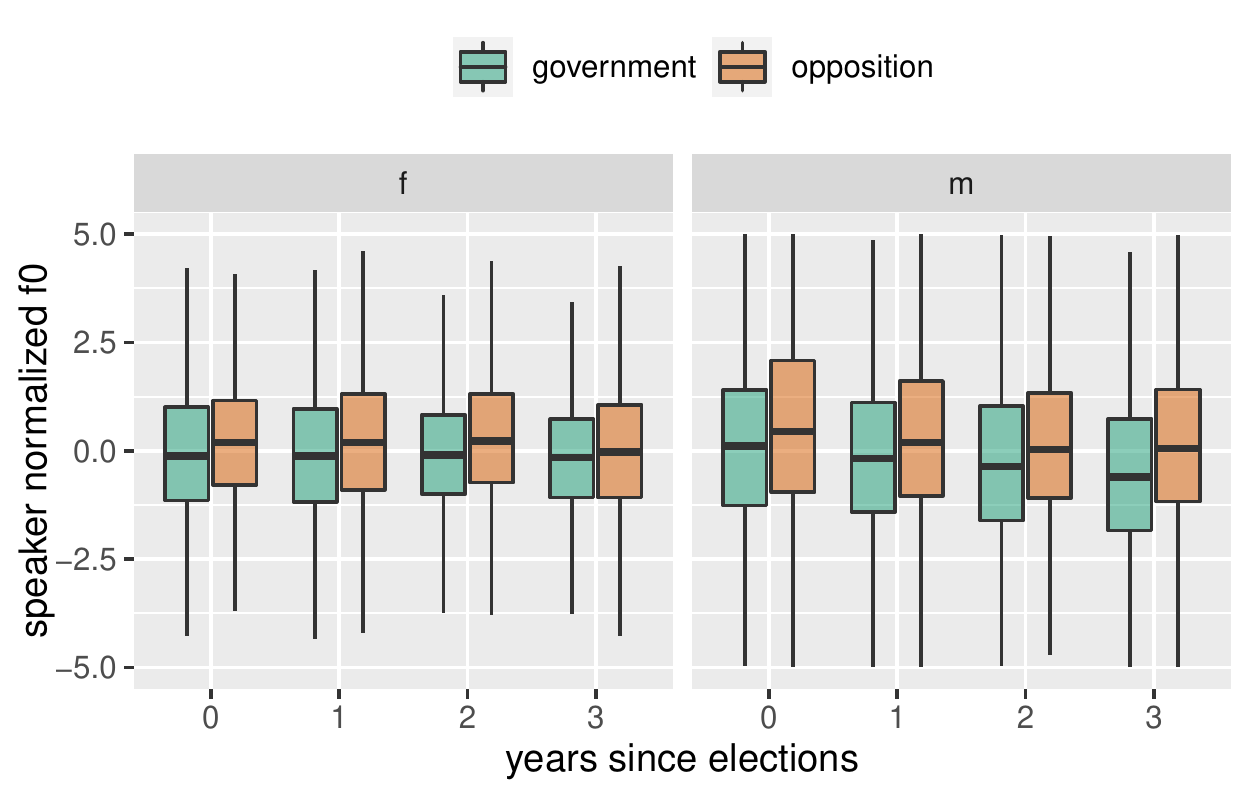}
  \caption{Normalized mean $f_0$ representing government and opposition throughout the election term.}
  \label{fig:since_election_sex_b}
\end{figure}

Finally, we have fitted a mixed effect model with the same dependent variable and with speaker's sex, political status, and the year count from the previous elections, with all interactions, as independent fixed effects and speaker's identity as a random factor (see Fig.~\ref{fig:since_election_sex_b} for the plot depicting the relationships). The only significant ($|t| > 2$) fixed effects were the following: sex, indicating slightly higher normalized $f_0$ for males ($t = 3.98$); opposition, confirming the higher $f_0$ for utterances spoken when the member represented opposition ($t=5.38$); and, importantly, the significantly negative years-since-election effect ($t = -5.70$) indicating a progressive $f_0$ decline over the election term. The only significant interaction was the one between male sex and election cycle ($t = -7.15$) showing that the decline in $f_0$ is steeper for males than for the females.

\section{Discussion}

The classification performance and the effect sizes based on the analysis of mean $f_0$ feature show subtle but robust differences between parliamentary utterances of government representatives and opposition. The aim of this work was, after all, not designing a reliable classification of individual utterances based on the political status (many utterances can be expected to be free of affect based influences) but to investigate potential status-based differences.  We have shown that the prosodic characteristics underlying these differences are manifested differently between sexes, and throughout the election term. We believe that other types of effects, such as the topical perceived increase in (affective) polarization \cite{lonnqvist2020polarization,bettarelli2022regional} can be studied using similar methods\footnote{The reason while our analysis does not reveal an increase in the distance between opposition and government representatives might be in the selection of the subset based on the long-term MPs.}.

While providing the statistical analysis for a single feature, it is known that $f_0$ 
 correlates with other features related to emotional arousal such as intensity \cite{titze1989relation}.  
 Moreover, simple summaries of pitch contours (e.g., mean or $f_0$
range) are sufficient to account for the most important variations observed between emotion
categories \cite{banziger2005role}. Such summaries have been successfully used in predicting United States Supreme Court judges' decisions
based on their oral arguments \cite{dietrich2019emotional}.

The corpus used in this study has been optimized for ASR-related tasks consisting of sentence fragments and lacks some important metadata. To study more nuanced phenomena would require full speeches to be recoverable from the corpus, and other information such as the type of speech and the topic of debate. Such information is available in a recently released semantic web portal ParlamenttiSampo \cite{hyvonen2023parliamentsampo} and could be used to augment the speech corpus used for a speech analysis.

\section{Acknowledgements}

\ifinterspeechfinal
     S.K. was supported by the Academy of Finland project no. 340125. The authors wish to acknowledge CSC – IT Center for Science, Finland, for providing the computational resources.
\else
     This will be added later.
\fi

\newpage

\bibliographystyle{IEEEtran}
\bibliography{mybib}

\end{document}